\documentclass[sn-mathphys]{sn-jnl}% Math and Physical Sciences Reference Style
\jyear{2022}%

\theoremstyle{thmstyleone}%
\theoremstyle{thmstyletwo}%

\theoremstyle{thmstylethree}%

\begin{document}

\title[Collisional- and photo-excitations]{Collisional- and photo-excitations of Ca IV 
including strong 3.2 $\mu$m emission line}


\author*[1]{\fnm{Sultana} \sur{Nahar}}\email{nahar.1@osu.edu}
\author[2]{\fnm{Bilal} \sur{Shafique}}\email{bilalshafiquekhawaja@ajku.edu.pk}
 \equalcont{These authors contributed equally to this work.}



\affil*[1]{\orgdiv{Astronomy Department}, \orgname{The Ohio State
University}, \orgaddress{\street{140 W. 18th Ave}, \city{Columbus}, 
\postcode{43210 }, \state{OH   }, \country{USA    }}}

\affil[2]{\orgdiv{Physics Department}, \orgname{University of Azad Jammu and Kashmir}, \orgaddress{\street{ }, \city{Muzzafarabad}, \postcode{13100}, 
\state{Kashmir }, \country{Pakistan}}}

%\affil[3]{\orgdiv{Department}, \orgname{Organization}, \orgaddress{\street{Street}, \city{City}, \postcode{610101}, \state{State}, \country{Country}}}


\abstract{
We report a detailed study of features of electron-impact excitation (EIE) of
Ca IV (Ca IV + e $\rightarrow$ Ca IV* + $e' \rightarrow$ Ca IV + h$\nu + 
e'$), for the first time using the relativistic Breit-Pauli R-Matrix method
with a large close coupling wavefunction expansion of 54 fine structure
levels belonging to n=2,3,4 complexes. Calcium lines in the infrared (IR) 
are expected to be observed by the high resolution James Webb Space Telescope.
Our study predicts presence of a strong 3.2 $\mu$m emission line in IR 
formed due to EIE of $3p^{5~ 2}P^o_{3/2} - 3p^{5~2}P^o_{1/2}$ in Ca IV. The 
EIE collision strength ($\Omega$) for the transition shows extensive 
resonances with enhanced background resulting in an effective 
collision strength ($\Upsilon$) of 2.2  at about 10$^4$ K that increases 
to 9.66 around 3$\times 10^5$ K. The present results include $\Omega$ of 
all 1431 excitation among the 54 levels and $\Upsilon$ for a limited number 
of transitions of possible interest. We have found extensive resonances in the 
low energy region of $\Omega$, convergence of the resonances and of the 
partial waves with the 54 levels wavefunction. 
At high energy $\Omega$ decreases beyond the resonance region for forbidden 
transitions, is almost constant or decreases slowly for dipole allowed 
transitions with low oscillator strengths (f-values), and rises with 
 Bethe-Coulomb behavior of ln(E) to almost a plateau for 
transitions with high f-values.
The wavefunction of Ca IV was obtained from optimization of 13 conﬁgurations
$3s^23p^5$, $3s3p^6$, $3s^23p^43d$, $3s^23p^44s$, $3s^23p^44p$,
$3s^23p^44d$, $3s^23p^44f$, $3s^23p^45s$, $3s3p^53d$, $3s3p^54s$, $3s3p^54p$,
$3p^63d$, $3s3p^43d^2$, each with the core configuration of $1s^22s^22p^6$,
using the atomic structure program SUPERSTRUCTURE. They produce 387 fine
structure levels. We report transition parameters - oscillator strengths, 
line strength ($S$) and $A$-values for a total of 93296 electric dipole 
(E1), quadrupole (E2), octupole (E3), magnetic dipole (M1) and quadrupole 
(M2) transitions among these levels. Lifetimes of these levels are also
presented.
%The abstract serves both as a general introduction to the topic and as 
%a brief, non-technical summary of the main results and their implications. 
%Authors are advised to check the author instructions for the journal 
%they are submitting to for word limits and if structural elements like 
%subheadings, citations, or equations are permitted.
}




\maketitle

\section{Introduction}\label{sec1}

Calcium is one of most abundant biogenic elements which is created
during supernova (SN) explosions. Lines of Ca I and II are seen in cool 
stars. The H and K lines and forbidden lines of Ca~II are commonly detected 
in SNe (.e.g. \cite{setal11}). The lines can determine the abundances of
the element in the object. Asplund et al \cite{aetal09} obtained Ca
abundance in the Sun with respect to hydrogen, in logarithmic abundances,
to be log[e(Ca)] = 6.34$\pm$0.04 where log[e(Ca])=12 + log[N(Ca)/N(H)], 
H is defined as log H = 12.00, N(Ca) and N(H) are the number densities of 
elements Ca and H, respectively. % log e(X) = 12+ log(NX/NH)+12
Ca-rich supernova, such as, SN 2019ehk in the star forming spiral galaxy 
Messier 100, about 55 million light years away from the earth, first by 
Shepherd \cite{js19} and analysis by Jacobson-Galan et al \cite{galenetal20}, 
brought much attention recently.
Jacobson-Galan \cite{galenetal20} found that the calcium-rich SN belongs to 
CaST class.% and is over abundant with almost a factor of 6
%in comparison to that in solar abundances \cite{getal17}. 
Ca-IV lines in infrared (IR) was observed in NGC 7027 nebula by
Feuchtgruber et al. \cite{fetal97}.  
3.2 $\mu$m line was observed in three Wolf-Rayet stars and was used to 
determine abundance and wind speed by Ignance et al \cite{ietal01}.
Furthermore, emission lines of
two calcium ions, Ca-V and Ca-VII, were observed in the NGC 2707 and
NGC 6302 nebulae \cite{fetal01}, and Ca~II lines in the high-temperature
environment of white dwarfs \cite{zetal03}.

The present work reports study of two atomic processes of Ca IV that
produce spectral lines, electron-impact excitation (EIE) and 
photo-excitation. EIE is one of the most common atomic process in near 
empty space or in astrophysical plasmas with or without a radiative source. 
A traveling electron interacts with an ion and transfers part of its energy 
and promotes the ion to an excited state. It is followed by the target 
ion de-exciting by emission of a photon, (e.g. \cite{aas11})
\begin{equation}
 { X^{+z} + e \rightarrow X^{+z*} + e' \rightarrow X^{+z} + h\nu + e' }
\end{equation}
where $X^{+z}$ is the target ion of charge $z$. The emitted photon can
form an observable line depending on the radiative transfer and density 
of the plasma. 
The projectile electron can also form a quasi-bound state by exciting the 
target ion to a higher energy state, $E^{**}$, while attaching itself to 
an orbit $\nu l$ which happens very commonly in nature, matching the
energy of a doubly excited autoionizing Rydberg state
\begin{equation}
E_x\nu l = E^{**}(X^{+z}\nu l) = E_x - z^2/\nu^2
\end{equation}
before going out free. This intermediate state introduces a resonance
in the collisional scattering. $E_x$ is the excited energy of the ion,
$\nu l$ are the effective quantum number and angular momentum of the
scattered electron. The present study implements close coupling (CC)
wavefunction expansion that produces the autoionizing resonances
automatically.

The other process, photo-excitation/ de-excitation, forms a line as the 
ion absorbs or emits a photon
\begin{equation}
 { X^{+z} + h\nu \leftrightarrow X^{+z*} }
\end{equation}
This process occurs most commonly when there is a radiative source, such 
as, a star shining the plasma.

Among the low ionization stages of Ca, Ca IV has been the least studied ion.
There is no study on electron impact excitation of it found in the
literature. EIE of Ca IV is the main focus of the present study.

Among the past studies on energies and transitions, Sugar and Corliss 
\cite{sc85} compiled the experimentally measured energy
levels of Ca IV which are available at NIST website \cite{nist}.
The radiative transition rates of Ca IV were reported by Naqvi \cite{n51},
Varsavsky \cite{v61}, Fawcett and Gabriel \cite{fg66}, Huang et al
\cite{h83}, Wilson et al. \cite{w00}, Gabriel et. al \cite{getal66} who
identified several lines generated from observed UV transitions.
The probability of detection of Ca IV lines has increased considerably
with high resolution observation of James Webb Space Telescope (JWST) in
the infra-red region.
The present work reports collisional excitation and photo-excitations for
many levels of Ca IV which include collision strengths ($\Omega$) and the
Maxwellian averaged collision strengths or effective collision strengths
($\Upsilon$), and parameters $f$-, $S-$ and $A-$ values for radiative 
transitions.
%The Introduction section, of referenced text \cite{bib1} expands on the background of the work (some overlap with the Abstract is acceptable). The introduction should not include subheadings.

%Springer Nature does not impose a strict layout as standard however authors are advised to check the individual requirements for the journal they are planning to submit to as there may be journal-level preferences. When preparing your text please also be aware that some stylistic choices are not supported in full text XML (publication version), including coloured font. These will not be replicated in the typeset article if it is accepted. 

\section{Theoretical Approximation}\label{sec2}

We give a brief outline of the theoretical background for electron impact
excitation and radiative photo-excitations below as guidance for the 
readers. More details can be found, e.g. in Pradhan and Nahar \cite{aas11}. 
We have treated EIE of Ca IV for collision strengths with relativistic 
Breit-Pauli R-matrix (BPRM) method, as developed under the Iron Project 
(IP, \cite{ip,betal95}). We used a wavefunction expansion in close-coupling 
(CC) approximation that includes excitation to n=2,3,4 levels in the target
and obtained collision strengths.  We obtained radiative transition
parameters for an extensive set of transition using relativistic Breit-Pauli 
approximation implemented in atomic structure program SUPERSTRUCTURE 
(SS, \cite{ejn74, necp03}).

Although two approaches are used for collision and photo-excitation, the
computations are related. BPRM calculations are initiated with the 
wavefunction expansion of the target ion, e.g. Ca IV, generated by 
program SUPERSTRUCTURE (SS). We discuss the outlines of collisional 
excitation first and then photo-excitations or radiative transitions.

\subsection{Breit-Pauli R-matrix (BPRM) calculations for EIE}

BPRM Hamiltonian as adopted under the Iron Project \cite{ip,betal95} in
atomic Rydberg unit is given by
\begin{equation}
%\begin{array}{l}
H_{N+1}^{\rm BP} =  \sum_{i=1}\sp{N+1}\left\{-\nabla_i\sp 2 -
\frac{2Z}{r_i}\right\} + \sum_{j>i}\sp{N+1} \frac{2}{r_{ij}}+
H_{N+1}^{\rm mass} +
H_{N+1}^{\rm Dar} + H_{N+1}^{\rm so}.
%+ i{Z\alpha^2 \over 4}\sum_i{\nabla^2({1 \over r_i})}, \\
%{\rm the~spin-orbit~interaction~term},~H^{\rm so}= Z\alpha^2
%\sum_i{1\over r_i^3} {\bf l_i.s_i}
%\end{array}
\end{equation}
where the first three terms belong to the non-relativistic Hamiltonian
and last three terms are the 1-body relativistic corrections which are
mass, Darwin, and spin-orbit interaction terms respectively. BPRM codes 
include all of them and part of the two-body correction terms of the
Breit-interaction (e.g. \cite{aas11}).
One Rydberg (Ry) is half of a Hartree giving the factor 2 in the terms.

BPRM calculations start with the target ion wavefunction generated by SS 
and calculates the wavefunction of the total atomic system of the target 
ion and 
the interacting electron in the close coupling (CC) approximation. In CC
approximation, the wavefunction of (e+ion) in a state $SL\pi J$, where $S$
is total spin, $L$ is the orbital, and $J$ is the total angular momenta,
is expressed as
\begin{equation}
 \Psi_E(e+ion) = A \sum_i^n \chi_i(ion)\theta_i + \sum_{j} c_j \Phi_j(e+ion)
\end{equation}
In the first term $\chi_i(ion)$ is the wavefunction expansion of the 
target ion, $\theta_i$ is that of the interacting electron, in channel
$S_tL_t\pi_tJ_tk_i^2l (SL\pi J)$ where $S_tL_t\pi_tJ_t$ is the target ion
state interacting with the projectile electron of energy $k_i^2$ and orbital
angular momentum $l$. The sum represents ground and various excited states
of the target ion. $A$ is the anti-symmetrization operator.
In the second term, $\Phi_j(e+ion)$ represents the (target+electron)
wavefunction, basically part of the first term separated out to show the
orthogonality condition of the interacting electron and short range 
interaction.
Close-coupling wave function expansion which includes target ion excitations 
enables producing the resonances inherently. The interference of the bound 
states of the target ion and the projectile electron continuum wavefunction
in the transition matrix introduces the resonances. 
Substitution of the CC expansion in the Schrodinger equation with the
Breit-Pauli Hamiltonian results in a set of coupled equations. The R-matrix
method is used to solve this set of equations for the energy and
wavefunctions of the (e+ion) system.

The scattering matrix for transition of the target ion from state $i$
to state $k$ by collision, ${\bf S}_{SL\pi J}(S_iL_iJ_il-S_kL_kJ_kl')$
where $SL\pi J$ is the (e+ion) state, $l$ and $l'$ are the incident and
scattered partial waves of the free electron, is derived from the reactant
matrix of the incident wave (e.g. \cite{br75,sb80,st82,aas11}).
The collision strength $\Omega$ for electron impact excitation (EIE) is
given by,
\begin{equation}
\Omega(S_iL_iJ_i-S_kL_kJ_k) =
 {1\over 2}\sum_{SL\pi J}
\sum_{l,l'}(2J+1) \lvert r {\bf S}_{SL\pi J}(S_iL_iJ_il - S_kL_kJ_kl')\rvert^2
\end{equation}
$\Omega$ reveals the detailed features with resonances of the collision.
The plasma models use the temperature dependent quantity, the effective
collision strengths $\Upsilon(T)$ which is obtained by averaging $\Omega$
over Maxwellian distribution function of the electrons at temperature $T_e$ as
\begin{equation}
\Upsilon_{ij}(T_e) = \int_0^{\infty} \Omega_{ij} (E)
\exp(-E/kT_e) d(E/kT_e),
\end{equation}
where $k$ is the Boltzmann constant and $E$ is the energy of the
projectile electron after the excitation, that is, the energy of the
scattered electron.
The excitation rate coefficient ($q_{ij}(T_e)$) is related to the effective 
collision strength $\Upsilon_{ij}$ as
\begin{equation}
q_{ij}(T_e) = \frac{8.63 \times 10^{-6}}{g_i T_e^{1/2}}
e^{-E_{ij}/kT_e} \Upsilon_{ij} (T_e) cm^3/s,
\end{equation}
where $g_i$ is the statistical weight of the initial level, T is in K,  
$E_{ij}$ is the transition energy in Rydberg, and (1/kT = 157885/T).

The high energy background of collision strength shows certain general
behaviors depending on the type of excitation. The background of
$\Omega$ is the smooth curve at the base of resonant features. For
forbidden transitions, $\Omega$ decreases to almost zero with higher 
energy.  For dipole allowed transitions,
%Bethe postulated the rising trend of $\Omega$ but with deceasing slope
%with higher energy.
using Born approximation with Coulombic wavefunction, $\Omega$ shows
a high energy limiting behavior and is given by Coulomb-Bethe approximation
\begin{equation}
\Omega_{ij}(E) = {4g_i f_{ij}\over E_{ij}}ln{E\over E_{ij}},
\end{equation}
where $f_{ij}$ is the oscillator strength for a dipole allowed transition,
and E is the incident electron energy. In the high energy limit,
\begin{equation}
\Omega_{ij}(E) \sim_{E\rightarrow \infty} d_{ij} ln(E)
\end{equation}
where $d_{ij}$ is proportional to the oscillator strength. The logarithmic
function will increase with increase in electron energy, but the the 
rising trend of the function slows down with very high values of the 
argument. Hence, for a low value of $d$, $\Omega$ is may not change as it 
is multiplied by a small number. For high value of $d$, $\Omega$ will 
increase but will lead toward a plateau.

\subsection{Atomic structure calculations for radiative transitions}

Theoretical details for obtaining radiative transition parameters
through atomic structure calculations using program SUPERSTRUCTURE can
be found, for example, in \cite{aas11,necp03}.
The Hamiltonian includes relativistic mass, Darwin, spin-orbit
interaction correction terms, full 2-body Breit interaction and
some additional two body terms.
The interacting electron and core ion potential, implemented in program
SS, is represented by Thomas-Fermi-Amaldi-Dirac potential. The program
uses configuration interaction wavefunction expansion, which for a
symmetry $J\pi$ can be expressed as
 \begin{equation}
 \Psi(\mbox{$J\pi$}) = \sum_{i=1}^N a_i\psi[C_i(\mbox{$J\pi$})]
 \end{equation}
where $a_i$s is the amplitude or the mixing coefficient of wavefunction
of configuration $C_i$, $\psi[C_i(\mbox{$J\pi$})]$ with symmetry $J\pi$,
and the sum is over all $N$ configurations that can produce
a level of symmetry $J\pi$.
The wavefunction will result in having $N$ number of eigenvalues from the
Hamiltonian matrix and each eigenvalue will correspond to energy of one
level of the symmetry. Accuracy of the energy of a level may depend on 
the size of the expansion and identification on the value of the mixing 
coefficient (e.g. \cite{aas11}.)

The transition matrix element, for example, for electric dipole allowed
transition (E1), is given by 
$<\Psi_B \lvert \lvert {\bf D} \rvert \rvert \Psi_{B'}>$ 
where $\Psi_B$ and $\Psi_{B'}$ are the initial and final state bound 
wavefunctions,
${\bf D}= \sum_i{\bf r}_i$ is the dipole operator where the sum is over
the number of electrons. The line strength $\bf S$ is obtained from the
mod squared of the transition matrix,
\begin{equation}
{\bf S= \lvert \left\langle{\mit\Psi}_f \vert\sum_{j=1}^{N+1} r_j\vert
 {\mit\Psi}_i\right\rangle\rvert^2 \label{eq:SLe} }
\end{equation}
where $\Psi_i$ and $\Psi_f$ are the initial and final wavefunctions. The
transition parameters, oscillator strength ($f_{ij}$) and radiative decay
rate ($A$) can be obtained from line strength as
\begin{equation}
%f_{ij} =  \left [{E_{ji}\over {3g_i}}\right ]S, ~~
 f_{ij} =  {E_{ji}\over {3g_i}} {\bf S}, ~~
%%\quad \sigma_{PI}(K\alpha,\nu)= 8.064 \frac{E_{ij}} {3g_{i}}
%%S^{\rm E1}~\mbox[Mb],
 A_{ji}(sec^{-1}) = \left [0.8032\times 10^{10}{E_{ji}^3\over {3g_j}}
 \right ]{\bf S}
\end{equation}
Transition probabilities for electric quadrupole (E2), magnetic dipole (M1),
electric octupole (E3), magnetic quadrupole (M2) transition parameters can
be obtained from their respective line strengths,(e.g. \cite{aas11,necp03}).

Lifetime of an excited level can be obtained from the inverse of the sum
of all transition probabilities to lower levels,
\begin{equation}
\tau_i(s) = 1/[\sum_j A_{ji}(s^{-1})]
\end{equation}
In atomic unit of time $\tau_0 = 2.4191\times 10^{-17}$s , the transition
probabilities or the radiative decay rate can be expressed as
$A_{ji}(s^{-1}) = {A_{ji}(a.u.)/\tau_0}$.

\section{Computation}\label{sec3}

The R-matrix calculations start with the target wavefunctions as an
input. These wavefunctions are obtained from atomic structure 
calculations, mainly using program SUPERSTRUCTURE (SS) \cite{ejn74,necp03}.
Ca IV wavefunction, energies and the relevant radiative transition parameters
were obtained from an optimized a set of 13 configurations of the ion,
 $3s^23p^5$, $3s3p^6$, $3s^23p^43d$, $3s^23p^44s$, $3s^23p^44p$, $3s^23p^44d$,
$3s^23p^44f$, $3s^23p^45s$, $3s3p^53d$, $3s3p^54s$, $3s3p^54p$, $3p^63d$,
$3s3p^43d^2$, with the same core configuration $1s^22s^22p^6$ for each,
using SS. The set of optimized Thomas-Fermi orbital scaling parameters 
are 1.26865(1s), 1.0395(2s), 1.04288(2p), 1.1(3s), 1.1(3p), 1.1(3d), 
1.083(4s), 1.079(4p), 1.031(4d), 1.1 (4f), 1.1 (5s) respectively.
The configurations set provided 387 fine structure levels, 54 of which
were considered for the collisional excitation in the present study and
hence used in the wavefunction expansion of Ca IV.
Selecting a set of levels of the target or core ion from a large set 
is the standard for a R-matrix calculation.  
Computations of collision strengths, which depend on the size of the
wavefunction expansion, needed several hundreds of CPU hours on 
the high performance computers at the Ohio Supercomputer Center. 

The purpose of having an optimized set of configurations, which produced a
large set of levels, is to ensure contributions from configuration 
interactions are included in the wavefunctions of levels,
and thus can provide a set of levels, starting from the ground level, of higher 
accuracy for the CC wavefunction expansion. 
Inclusion of all levels, 387 in the present case, in the calculations will
require considerably large computational time, and expected to be 
computationally prohibitive which is very common, and hence impractical 
to consider. The set of excited levels 
considered for a R-matrix calculation is typically based on the expected 
physics to be revealed, mainly the resonant features. No new physics was 
expected from levels higher than 54 levels since the resonances would have 
converged and hence those belonging to higher ones would no longer be strong 
but have weakened to converge to the background. This was also our finding
as demonstrated in the Results section.

Table~1 presents the 54 fine structure levels of Ca IV included in the
wavefunction expansion and compares the calculated energies from SS with
experimental values tabulated by the National Institute of Standards and
Technology (NIST) (www.nist.gov). The comparison shows that the present 
computed energies are generally within a few percent of the observed values.
The accuracy of a level energy depends on the how well the wavefunction
expansion is representing the level through interaction of the given set
of configurations.
For more precise energy positions of the resonances in EIE collision 
strength, we have replaced the calculated energies with the available
observed energies in the BPRM calculations.
\begin{table}
\caption{\label{tabl1}Comparison of the present calculated energies for
the 54 fine structure levels of Ca IV with those (Sugar and Corliss
\cite{sc85}) available in the compilation table of the NIST \cite{nist}}
\footnotesize
\begin{tabular}{@{}lllllc}
\toprule
K &Configuration&Term&E$_{present}$&E$_{NIST}$ & \% diff\\
 & & & (Ry) & (Ry) & \\
\midrule
1 & $3s^23p^5$ & $^2P^o_{3/2}$ & 0 & 0 & 0.0 \\
2 & $3s^23p^5$ & $^2P^o_{1/2}$ & 0.0302   & 0.0284 & 5.9 \\
3 & $3s3p^6$ & $^2S_{1/2}$     & 1.3031   & 1.3891 & 6.2 \\
4 & $3s^23p^43d$ & $^4D_{7/2}$ & 1.8197   & 1.8362 & 0.9 \\
5 & $3s^23p^43d$ & $^4D_{5/2}$ & 1.8218   & 1.8385 & 0.9 \\
6 & $3s^23p^43d$ & $^4D_{3/2}$ & 1.8243   & 1.8410 & 0.9 \\
7 & $3s^23p^43d$ & $^4D_{1/2}$ & 1.8263   & 1.8431 & 0.9\\
8 & $3s^23p^43d$ & $^4F_{9/2}$ & 1.9901   & 1.9865 & 0.2 \\
9 & $3s^23p^43d$ & $^4F_{7/2}$ & 1.9999   & 1.9964 & 0.2  \\
10 & $3s^23p^43d$ & $^2P_{3/2}$ & 2.0047. & 1.9965 & 0.4 \\
11 & $3s^23p^43d$ & $^2P_{1/2}$ & 2.0070  & 2.0036 & 0.2 \\
12 & $3s^23p^43d$ & $^4F_{5/2}$ & 2.0115  & 2.0083 & 0.2 \\
13 & $3s^23p^43d$ & $^4F_{3/2}$ & 2.0225. & 2.0149 & 0.4 \\
14 & $3s^23p^43d$ & $^4P_{5/2}$ & 2.0705. & 2.0473 & 0.2  \\
15 & $3s^23p^43d$ & $^4P_{3/2}$ & 2.0761  & 2.0536 & 1.1 \\
16 & $3s^23p^43d$ & $^2D_{3/2}$ & 2.0817  & 2.0732 & 0.4 \\
17 & $3s^23p^43d$ & $^4P_{1/2}$ & 2.0840  & 2.0616 & 1.0 \\
18 & $3s^23p^43d$ & $^2D_{5/2}$ & 2.0970  & 2.0890 & 0.4 \\
19 & $3s^23p^43d$ & $^2F_{7/2}$ & 2.1076  & 2.1074 & 0.02\\
20 & $3s^23p^43d$ & $^2F_{5/2}$ & 2.1310  & 2.1297 & 0.06 \\
21 & $3s^23p^43d$ & $^2G_{7/2}$ & 2.1369  & 2.1573 & 0.95 \\
22 & $3s^23p^43d$ & $^2G_{9/2}$ & 2.1382  & 2.1571 & 0.88  \\
23 & $3s^23p^43d$ & $^2F_{7/2}$ & 2.2990  & 2.3051 & 0.26 \\
24 & $3s^23p^43d$ & $^2F_{5/2}$ & 2.3046  & 2.3109 & 0.27\\
25 & $3s^23p^43d$ & $^2D_{5/2}$ & 2.4602  & 2.4937 & 1.36\\
26 & $3s^23p^43d$ & $^2D_{3/2}$ & 2.4653  & 2.4993 & 1.34 \\
27 & $3s^23p^43d$ & $^2S_{1/2}$ & 2.6559  & 2.6979 & 1.8 \\
28 & $3s^23p^44s$ & $^4P_{5/2}$ & 2.6688  & 2.7714 & 3.7 \\
29 & $3s^23p^44s$ & $^4P_{1/2}$ & 2.6701  & 2.7972 & 4.5  \\
30 & $3s^23p^44s$ & $^2P_{3/2}$ & 2.6780  & 2.7974 & 4.24  \\
31 & $3s^23p^44s$ & $^4P_{3/2}$ & 2.6818  & 2.7857 & 3.6  \\
32 & $3s^23p^44s$ & $^2P_{1/2}$ & 2.6895  & 2.8098 & 4.3  \\
33 & $3s^23p^43d$ & $^2D_{5/2}$ & 2.7449  & 2.8566 & 4.2 \\
34 & $3s^23p^43d$ & $^2P_{3/2}$ & 2.7495  & 2.8976 & 5.2  \\
35 & $3s^23p^43d$ & $^2P_{1/2}$ & 2.7666  & 2.9135 & 4.9  \\
36 & $3s^23p^44s$ & $^2D_{3/2}$ & 2.7689  & 2.8794 & 3.8  \\
37 & $3s^23p^44s$ & $^2D_{5/2}$ & 2.8479  & 2.9941 & 4.7  \\
38 & $3s^23p^43d$ & $^2D_{3/2}$ & 2.8491  & 2.9961 & 4.7  \\
39 & $3s^23p^44p$ & $^4P^o_{5/2}$ & 3.0145 & 3.1268& 3.8   \\
40 & $3s^23p^44p$ & $^4P^o_{3/2}$ & 3.0179 & 3.1318& 3.5   \\
41 & $3s^23p^44p$ & $^4P^o_{1/2}$ & 3.0251 & 3.1386& 3.5   \\
42 & $3s^23p^44p$ & $^4D^o_{7/2}$ & 3.0539 & 3.1829& 4.1   \\
43 & $3s^23p^44p$ & $^4D^o_{5/2}$ & 3.0610 & 3.1913& 4.0   \\
44 & $3s^23p^44p$ & $^4D^o_{3/2}$ & 3.0706 & 3.1996& 4.0   \\
45 & $3s^23p^44p$ & $^2P^o_{1/2}$ & 3.0729 & 3.2128& 4.3   \\
46 & $3s^23p^44p$ & $^4D^o_{1/2}$ & 3.0751 & 3.2039& 4.0   \\
47 & $3s^23p^44p$ & $^2D^o_{5/2}$ & 3.0824 & 3.2185& 4.0   \\
48 & $3s^23p^44s$ & $^2S_{1/2}$   & 3.0828 & 3.2573& 5.4   \\
49 & $3s^23p^44p$ & $^2D^o_{3/2}$ & 3.0888 & 3.2376& 4.0   \\
50 & $3s^23p^44p$ & $^2P^o_{3/2}$ & 3.1017 & 3.2209& 3.7   \\
51 & $3s^23p^44p$ & $^4S^o_{3/2}$ & 3.1217 & 3.2577& 4.0   \\
52 & $3s^23p^44p$ & $^2S^o_{1/2}$ & 3.1249 & 3.2598& 4.0   \\
53 & $3s^23p^44p$ & $^2F^o_{7/2}$ & 3.2091 & 3.3629& 4.5   \\
54 & $3s^23p^44p$ & $^2F^o_{5/2}$ & 3.2133 & 3.3591& 4.5   \\
\botrule
\end{tabular}
\end{table}

It is important to ensure convergence in contributions by the number of
partial waves and (e+ion) symmetries to the collision strengths $\Omega$.
We computed $\Omega$ several times by varying different sets of partial 
waves $l$ going up to 22 and (e-ion) symmetries J going up to $\leq 22$ of 
even and odd parities. 
We found that i) $\Omega$ background is converged with the highest values
of $l$ = 20 and $J\pi$ = 11, ii) larger number of $l$ and $J\pi$ than these
introduced computational instability which gives NaN (Not a Number) for 
the collision strengths at various electron energies. 

Getting "NaN" for $\Omega$ values is a known problem for a R-matrix calculation. 
They are introduced when the computation goes through very small numbers.
Usually the NaN points are deleted from the sets of $\Omega$ values. 
However, we made attempts to get approximate values $\Omega$ at energies of
NaN values obtained with large $l$ and $J$ values by extrapolating or 
interpolating the neighboring $\Omega$ points. Smooth background can be 
extrapolated but
resonances are not reproduced in $\Omega$ as can be seen in the blue (with 
extrapolation) and red curves (no extrapolation) in Figure 1 representing 
the same excitation of the ion. 
The resonant peaks of the blue curve, which has many NaN points and 
interpolation was carried out to replace the NaNs with real numbers, are 
lower than the red curves. 

Figure 1 also demonstrates the test of convergence of $\Omega$ for a few 
sets of highest values of $l$ and $J$ for the excitations of a) $^2P^o_{3/2} - 
^2P^o_{1/2}$ and b) $^2P^o_{1/2} - ^2S_{1/2}$ computed at a coarse energy 
mesh. The red curves 
correspond to use of $l$ = 0 - 20 and $J\pi$ = 0- 11, the blue ones to 
$l$ = 0 - 22 and $J\pi$ = 0 - 21 and the magenta ones to $l$ = 0 - 11 and 
$J\pi$ = 0 - 10. 
 $\Omega$ in magenta ($l$ = 0- 11 and $J\pi$= 0- 10) are considerably 
lower than the other curves indicating that convergence of contributions
from 12 partial waves has not been reached. The blue and the red 
curves have about the same background indicating convergence in contributions
of partial waves has reached. 
Hence specifications for the red curve, $l$ = 0 - 20 and $J\pi$ = 0- 11, 
have been used for the computation of $\Omega$.
Final $\Omega$ values were obtained at fine energy meshes. We used a
very fine energy mesh of $\Delta E < 10^{-6}$ Rydberg to resolve the
near-threshold resonances.
 \begin{figure}
\vskip -0.25in
%\hskip -0.3in
  \includegraphics[height=4.3in,width=5.0in]{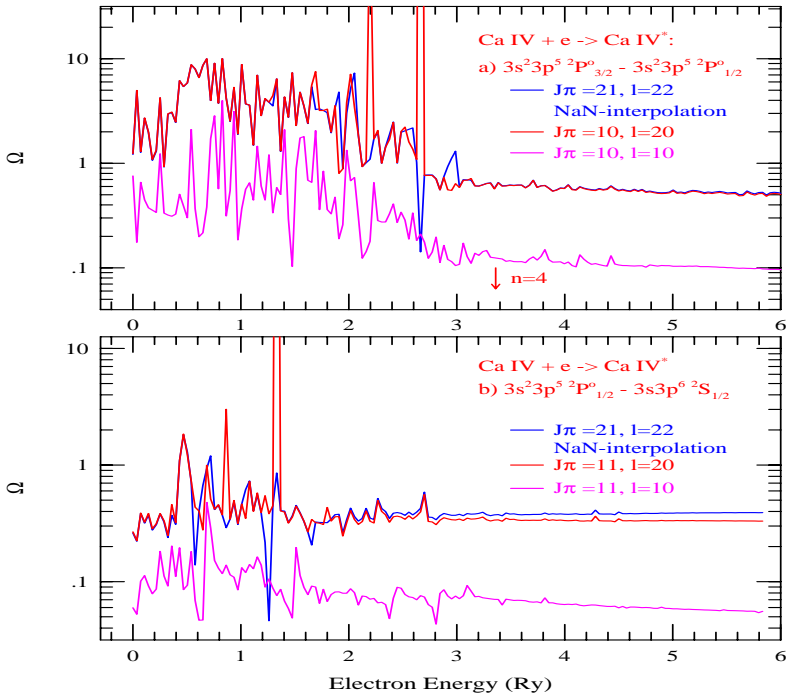}
 \caption{Demonstration of convergence of contributions of partial
waves with various $l$ and $J$ values to $\Omega$. The x-axis
corresponds to the energy of the scattered electron after the
excitation which starts with zero energy. Red curves indicate the
best convergent condition.}
 \end{figure}

The above discussion concerns impact of partial waves that are 
noticeable. Beyond it, top-up which gives some additional contributions 
is added. As top-up contributions, we included contributions of higher 
multipole potentials to $\Omega$ using the option ipert=1 
in STGF of the R-matrix codes. The other top-up contribution 
can come from higher partial waves, beyond the specified ones, using the 
option chosen as ipert=2, in STGF. The approximation incorporated in the
R-matrix code STGF is most probably based on the treatment of Burgess and 
Tully \cite{bt78} of higher partial waves. 
However, computation of contributions of the higher partial waves, 
not only takes much longer time, it is not done for most cases as it stops 
due to numerical issues except for a few excitations and at some energies. 
We computed contributions of higher partial waves when it is possible to 
test them and found negligible contributions in increasing the $\Omega$ 
values to affect the average collision strengths $\Upsilon$. The problem 
is compensated, as done in the present case, by including larger number 
of partial waves without the approximation. 

Program "ecs-omg.f" \cite{snn13} was used to calculate the effective 
collision strengths 
$\Upsilon$, Eq.(7), at various temperatures where $\Omega$
is integrated over the energy of the scattered electron from zero to
a high value. The high energy limit is chosen to a value at which $\Omega$ 
has diminished to a near zero value or has reached a plateau and the 
exponential factor of $\Upsilon$ has approached a near zero value. $\Omega$ 
points between the highest electron energy computed by the BPRM codes to 
the highest energy limit of the $\Upsilon$ integral are obtained using 
the logarithmic behavior of Coulomb-Bethe approximation of Eq.(10). 

The radiative data of $f-$ and $A-$values for dipole allowed photo-excitations 
(E1) have been reprocessed with experimental energies using code PRCSS 
(e.g. \cite{snn14}). This allows to obtain the transition parameters at
observed wavelengths. For the reprocessing, the transition energies were 
obtained from the experimental level energies and then multiplied, 
following Eq. 13.  to the calculated line strengths from code SUPERSTRUCTURE.
For the levels for which no observed or measured values are available, 
calculated energies were used.

\section{Results and Discussions}\label{sec4}

We present atomic parameters for electron impact excitation of
(e + Ca IV $\rightarrow$ Ca IV$^*$ + e' $\rightarrow$ Ca IV + h$\nu$
+ e') and photo-excitation of (Ca IV + h$\nu~ \leftrightarrow$ Ca IV$^*$).
The results for the collisional excitation are reported for the first
time, as indicated by literature search. They are described below
first followed by those for photoexcitation.

\subsection{Collisional excitation of Ca IV}

We discuss the characteristic features of collisional excitation with
illustrative examples. The first excitation of the target is usually of
particular interest because of its high probability through EIE and 
the emitted photon, typically of low energy, can travel for a long time 
without being absorbed. If the corresponding emission line is strong, 
it can be detected easily in low density plasmas and be used 
for identification of the ion and environmental diagnostics.
The first excitation in Ca IV, $3p^5~^2P^o_{3/2} - 3p^5~^2P^o_{1/2}$,
within the ground configuration is of particular importance since the
wavelength of the emission line, 3.207 $\mu$m, is well within the high 
resolution IR wavelength detection range, 0.6 - 28.3 $\mu$m, of JWST, 
and could be used for diagnostics, abundances (e.g. \cite{ietal01}).

Figure 2 upper panel presents collision strength for the first excitation
$\Omega(^2P^o_{3/2} - ^2P^o_{1/2})$ of the target ion Ca IV with respect
to scattered electron energy after the excitation. The electron energy is
relative to the excitation threshold and hence starts at zero.
We note that the $\Omega$ for the collisional excitation is quite strong
as it shows extensive resonances with enhanced background in the energy
region between the first excited level $^2P^o_{1/2}$ and the next one
$^2S_{1/2}$ (pointed by arrows), and continues to be strong beyond
it up to $^2D_{3/2}$ level, the one before the last dipole allowed
transition in the 54 level wavefunction expansion.
Beyond, $^2D_{3/2}$ the transitions are forbidden except one and the
resonances become weaker.
There are 29 dipole allowed levels in total that exist in this energy 
range.  Each excitation of the target ion corresponds to a Rydberg 
series of resonances.
Typically, the resonances corresponding to a dipole allowed excitation are
visible while others are suppressed. The resonances belonging to very high
energy levels are found to become weaker. Such weakening trend indicates 
convergence of resonant contributions to the collisional parameters.

Figure 2 lower panel presents effective collision strength ($\Upsilon$) 
for $\Omega[(^2P^o_{3/2} - ^2P^o_{1/2})$. Starting low at lower energy, 
$\Upsilon$ forms a shoulder bump at 10$^4$ K. Then it rises
relatively quickly reaching to the high peak value 9.66 at about 3$\times
10^5$ K. The peak indicates existence of a strong line of the transition 
in Ca IV which can be detected. The intensity of the line will depend on
the plasma environment.
  \begin{figure}
  \vskip -0.35in
  \hskip -0.50in
\centering
  \includegraphics[height=7.0in,width=5.1in]{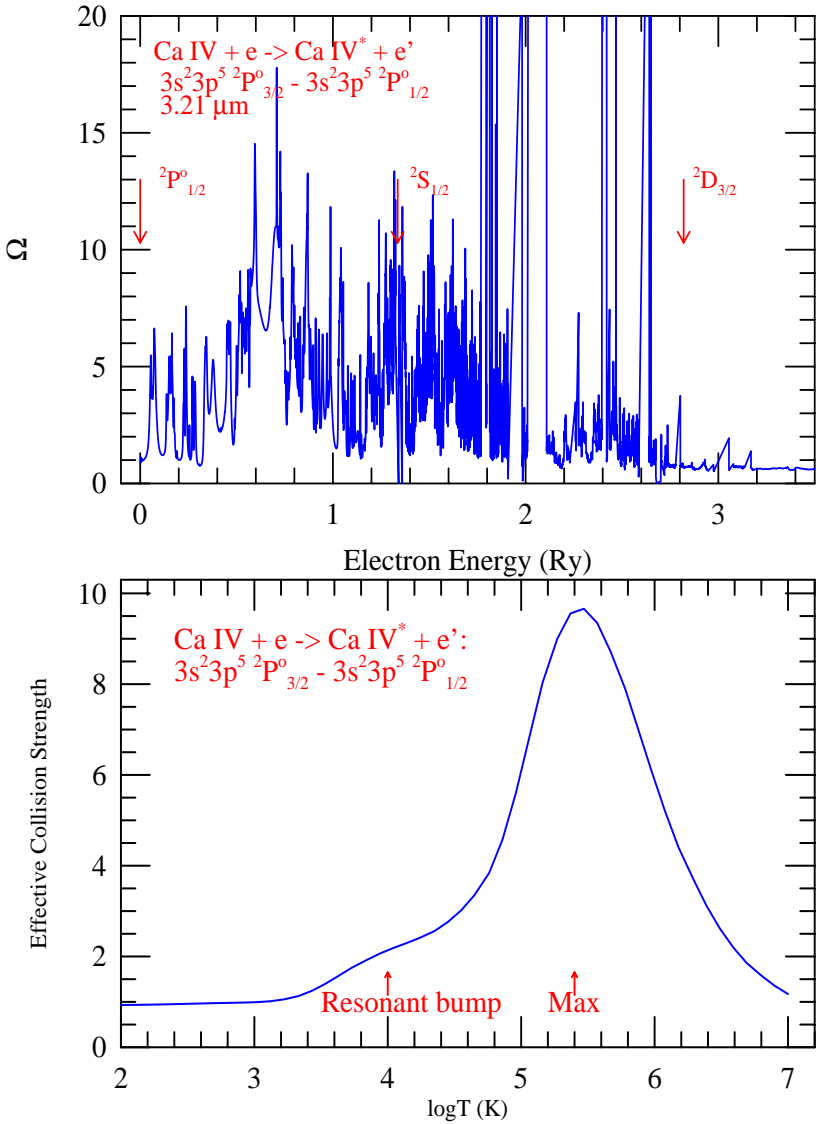}
  \vskip -0.20in
  \caption{Upper panel: EIE collision strength ($\Omega$) of Ca IV for 
 the first excitation $3s^23p^5~^2P^o_{3/2} - 3s^23p^5~^2P^o_{1/2}$ of 
 the ground level with respect to scattered electron energy in Ry unit.
 Extensive resonances with enhanced background can be noted within the 
 energy region of dipole allowed transitions from the ground level 
 $3s^23p^5~^2P^o_{3/2}$ up to $^2D^o_{1/2}$.
 Lower panel: Effective collision strength for $3s^23p^5~^2P^o_{3/2} - 
 3s^23p^5~^2P^o_{1/2}$ excitation in Ca IV showing a high peak value
 at about 3$\times 10^5$ K indicating high probability of detection 
of the line by JWST.
 }
  \end{figure}

% $\Omega$ for the forbidden transitions shows resonances with low
%background which diminishes almost to zero at higher energy. 
We present illustrative examples of forbidden transitions in Ca IV.
Figure 3 presents collision strengths for two forbidden excitations in
Ca IV, a) $\Omega(3s^23p^5~^2P^o_{3/2} - 3s^23p^44s ^4D_{7/2})$ and b)
$\Omega(3s^23p^5~^2P^o_{3/2} - 3s^23p^44s ^4F_{9/2})$. Both transitions show
presence of strong resonances in the lower energy region. The resonances
become  weaker in higher energy region indicating contribution of
resonances is converging. This is the typical trend of
$\Omega$ for forbidden transitions. Both of these transitions lie in the
extreme ultraviolet region
with wavelengths of 496 and 458 $\AA$ respectively.
 \begin{figure}
 \vskip -0.15in
 \hskip -0.25in
  \includegraphics[height=4.5in,width=5.2in]{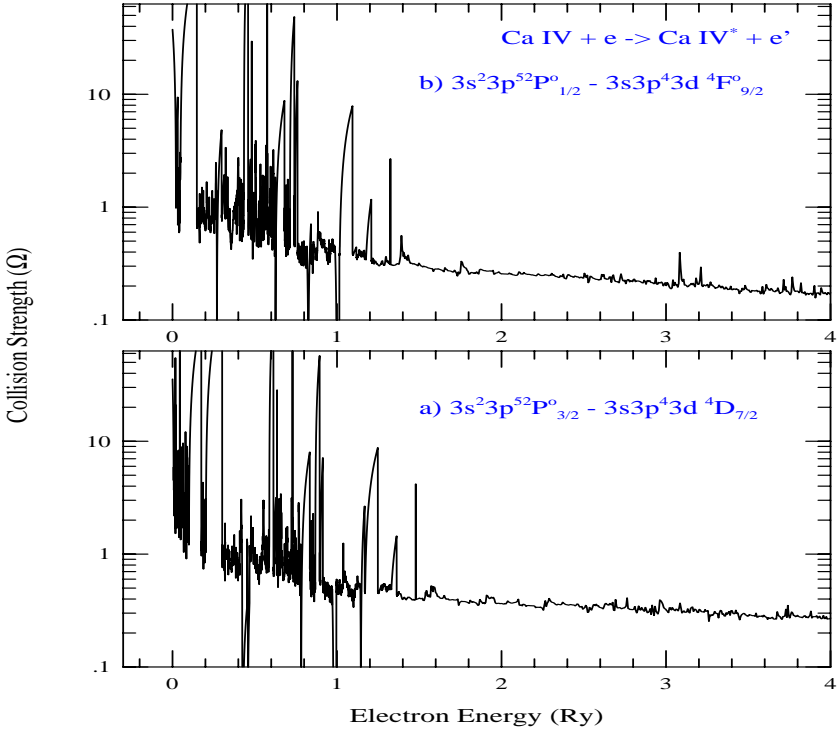}
  \caption{Collision strength for forbidden excitation
 a) $\Omega(3s^23p^5~^2P^o_{3/2} - 3s^23p^44s~^4D_{7/2})$ and
 b) $\Omega(3s^23p^5~^2P^o_{3/2} - 3s^23p^44s~^4F_{9/2})$ in extreme
 ultraviolet, 496 $\AA$ and 458 $\AA$ respectively, illustrating resonant 
 features in the low energy region and decreasing background in the high  
 energy region.}
  \end{figure}

 Collision strengths for dipole allowed transitions may show a different 
 trend in the high energy background from those of forbidden
 transitions. The dipole in the target can affect the partial waves of
 the incident electron and contribute to the collision strength. The
 contribution depends on the oscillator strength for the dipole transition.
 For stronger transitions, inclusion of larger number of partial waves is 
 important for converged contributions.

 \begin{figure}
 \vskip -0.15in
 \hskip -0.2in
 \includegraphics[height=4.5in,width=5.1in]{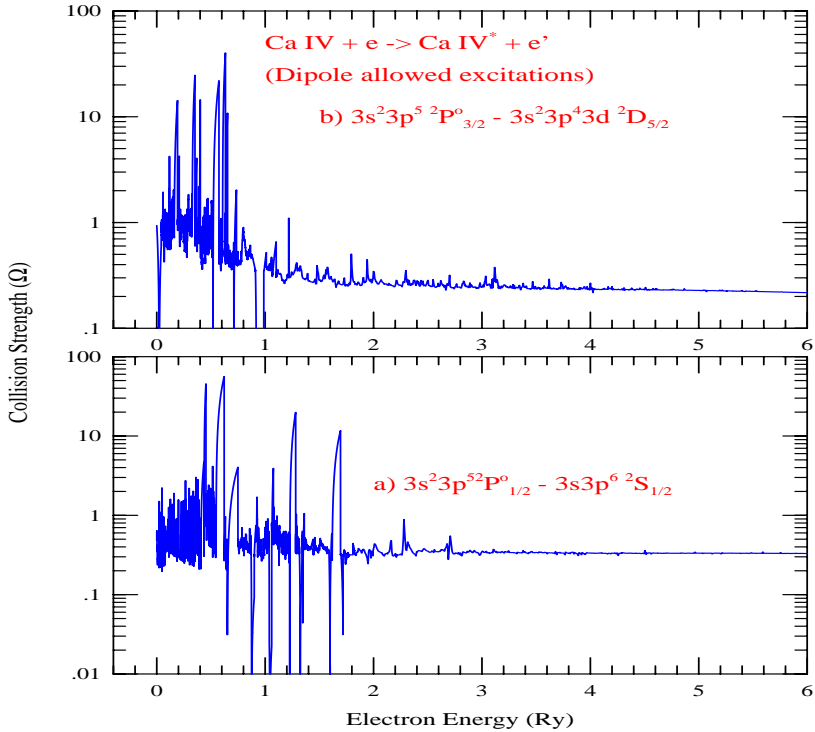}
 \caption{Features of $\Omega(EIE)$ for two weaker dipole allowed transitions
(oscillator strengths are given in Table 2): a) $3s^23p^5~^2P^o_{1/2} - 
3s3p^6~^2S_{1/2}$ (levels 2-3) and b) relatively higher transition 
 $3s^23p^5~^2P^o_{3/2} - 3s^23p^4 3d ^2D_{5/2}$ (levels 1 -18). Resonances 
have converged to a smooth background at high energy where the background 
remains almost constant or decreases slowly with increase of energy, typical
for weak transitions.
}
 \end{figure}
Figure 4 presents features of $\Omega$ (EIE) for two dipole allowed
transitions to low lying excited levels, a) $3s^23p^5~^2P^o_{1/2} - 
3s3p^6~^2S_{1/2}$ (levels 2-3) at 670 $\AA$ and b) $3s^23p^5~^2P^o_{3/2} -
3p^43d~  ^2D_{5/2}$ (levels 1-18)
at 435 $\AA$. $\Omega$ for these transitions show presence of prominent
resonances in the lower energy regions which become weaker converging to 
smooth background at higher energy. 
It can be noted that the high energy background is decreasing very slowly 
or remaining almost constant.
These transitions are much weaker compared to others as indicated by their
smaller values for the oscillator strengths and $A$-values presented in 
Table 2.
  \begin{figure}
   \vskip -0.25in
   \hskip -0.2in
   \includegraphics[height=4.5in,width=5.0in]{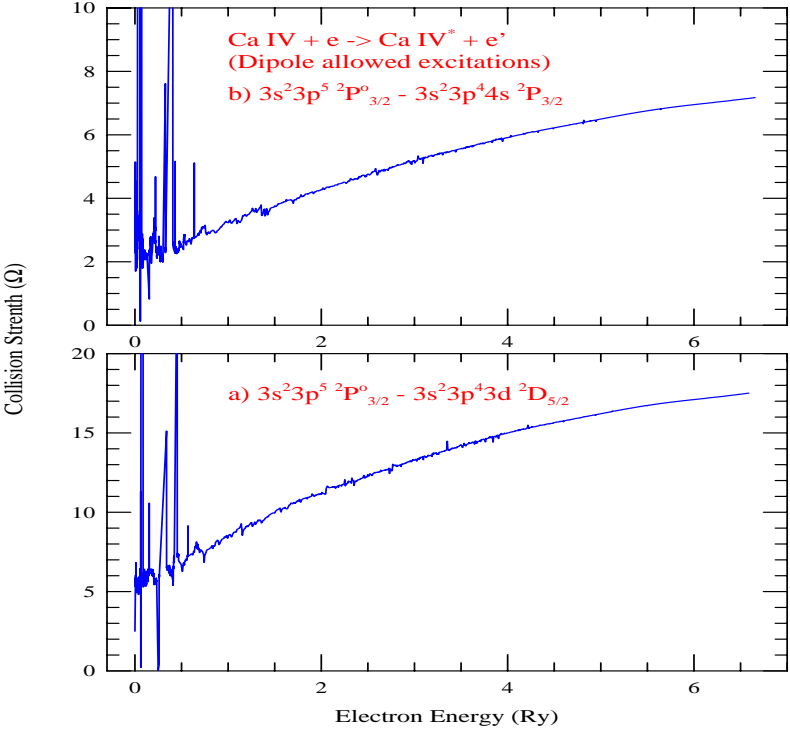}
  \caption{Features of $\Omega$ for excitation to high lying dipole allowed
 levels, a)  $3s^23p^5~^2P^o_{3/2} - (3s^23p^43d)^2D_{5/2}$ and  b)
$3s^23p^5~^2P^o_{3/2} - (3s^23p^44s)~^2P_{3/2}$ with stronger or larger
values for oscillator strengths illustrating (a) convergence of resonances 
and (b) rising Coulomb-Bethe $ln(E)$ behavior of $\Omega$ at higher energy. 
The background of $\Omega$ rises toward forming a plateau.
 %$^2P^o_{3/2} - (3s^23p^43d)^2S_{1/2}$ and  b) $^2P^o_{3/2} - 
 %(3s^23p^44s)^2P_{1/2}$.
  }
  \end{figure}

 \begin{table}
 \caption{Dipole allowed transitions between levels $i$ and $j$, their $f$
and A-values, and transition wavelength $\lambda_{ij}$ in Angstrom unit for
the illustrated examples of $\Omega$ of Ca IV in Figures 4 and 5. The level
indices $i$ and $j$ correspond to those in Table 1.}
 %\footnotesize 
%\begin{tabular}{@{}lllllcc} 
 \begin{tabular}{lllllcc}
 \toprule
 Figure&\multicolumn{1}{c}{i - j} & \multicolumn{2}{c}{$SL\pi J$} &
 $\lambda_{ij}$ & $f_{ij}$ & $A_{ji}$(s$^{-1)})$\\
 & &  $i$ & $j$ & ($\AA$) & & \\
 \midrule
 4a & 2 - 3&$^2P^o_{1/2}$& $^2S_{1/2}$ &   670 & 1.55E-02 & 2.01E+08\\
 4b & 1 -18&$^2P^o_{3/2}$& $^2D_{5/2}$ &   435 & 8.14E-03 & 1.98E+08\\
 5a & 1 -33&$^2P^o_{3/2}$& $^2D_{5/2}$ &   319 & 2.60E+00 & 1.14E+11\\
 5b & 1 -30&$^2P^o_{3/2}$& $^2P_{3/2}$ &   332 & 1.09E+00 & 7.10E+10\\
\botrule
\end{tabular}
\end{table}

% \begin{figure}
%  \vskip -0.35in
%  \hskip -0.2in
%  \includegraphics[height=4.5in,width=5.0in]{omgca4-130-133}
%%  \includegraphics[height=4.5in,width=5.0in]{fig5.eps}
% \caption{Features of $\Omega$ for excitation to high lying dipole allowed
%levels, a)  $3p^5~^2P^o_{3/2} - (3s^23p^43d)^2D_{5/2}$ and  b)
%$3p^5~^2P^o_{3/2} - (3s^23p^44s)~^2P_{3/2}$ with stronger oscillator
%strengths illustrating (a) convergence of resonances and (b) rising
%Coulomb-Bethe $ln(E)$ behavior of $\Omega$ at higher energy. The
%background of $\Omega$ rises until tending to form a plateau.
% }
% \end{figure}
Figure 5 presents $\Omega$ for dipole allowed excitation of
a) $3s^23p^5~^2P^o_{3/2} - (3s^23p^43d)^2D_{5/2}$ at 340 $\AA$ and b)
$3s^23p^5~^2P^o_{3/2} - (3s^23p^44s)~^2P_{3/2}$ at 332 $\AA$ in Ca IV.
The high peak resonances are converging near 1 Ry. Although there
are much more core ion excitations included in the wavefunction expansion, 
the Rydberg series of resonances belonging to them have become weak 
with almost no contributions to $\Omega$. This indicates convergence 
in the wavefunction expansion for generating resonant features. However, 
the background of $\Omega$ is rising with energy toward forming a plateau. 
This trend is in agreement with the expected Coulomb-Bethe behavior of 
$\Omega_{if}(E) \sim_{E\rightarrow \infty} d ln(E)$ (Eq.10) at high energy
for transitions with stronger or larger values for oscillator strength or 
$A$-values. The $f$- and $A$-values for for the two transitions are given 
in Table 2.
As discussed above, for these transitions, larger number of partial waves
needs to be considered in calculating $\Omega$. The rising trend toward a
plateau may not result in with inadequate number of partial waves. In 
such case, $\Omega$ is often extrapolated with Coulomb-Bethe form. 

Astrophysical models require temperature dependent effective collision
strength ($\Upsilon$). Table 3 presents $\Upsilon$ for some excitations to
levels that could be of importance for astrophysical applications.
$\Upsilon$ for any other excitation of the 1431 transitions can be
computed by averaging $\Omega$ over Maxwellian distribution function
at any temperature. The $\Omega$ values, and $\Upsilon$ for a number of
excitations, in addition to the ones presented here, are available 
electronically at NORAD-Atomic-Data \cite{norad}.
\begin{table}
\caption{\label{tabl2} $\Upsilon$ at various temperatures for 
different excitations of levels 1-2 to 1-18 in Ca IV.
}
\footnotesize
\begin{tabular}{lllllll}
\toprule
 logT& T (K)& $\Upsilon$:  1- 2& 1- 3& 2- 3& 1- 4& 1- 5\\
\midrule
 2.00& 1.000E+02& 9.312E-01& 1.038E+00& 5.709E-01& 3.274E+01& 2.158E-01\\
 2.10& 1.265E+02& 9.339E-01& 1.061E+00& 5.674E-01& 3.199E+01& 2.395E-01\\
 2.20& 1.600E+02& 9.382E-01& 1.075E+00& 5.587E-01& 3.104E+01& 2.827E-01\\
 2.31& 2.024E+02& 9.439E-01& 1.077E+00& 5.447E-01& 2.986E+01& 3.545E-01\\
 2.41& 2.560E+02& 9.507E-01& 1.065E+00& 5.261E-01& 2.841E+01& 4.608E-01\\
 2.51& 3.237E+02& 9.581E-01& 1.040E+00& 5.037E-01& 2.669E+01& 6.014E-01\\
 2.61& 4.095E+02& 9.654E-01& 1.002E+00& 4.791E-01& 2.474E+01& 7.688E-01\\
 2.71& 5.179E+02& 9.722E-01& 9.568E-01& 4.537E-01& 2.261E+01& 9.507E-01\\
 2.82& 6.551E+02& 9.785E-01& 9.081E-01& 4.294E-01& 2.042E+01& 1.134E+00\\
 2.92& 8.286E+02& 9.851E-01& 8.608E-01& 4.081E-01& 1.826E+01& 1.315E+00\\
 3.02& 1.048E+03& 9.948E-01& 8.192E-01& 3.909E-01& 1.623E+01& 1.503E+00\\
 3.12& 1.326E+03& 1.014E+00& 7.854E-01& 3.784E-01& 1.440E+01& 1.724E+00\\
 3.22& 1.677E+03& 1.054E+00& 7.604E-01& 3.702E-01& 1.284E+01& 2.002E+00\\
 3.33& 2.121E+03& 1.127E+00& 7.436E-01& 3.654E-01& 1.156E+01& 2.335E+00\\
 3.43& 2.683E+03& 1.242E+00& 7.345E-01& 3.634E-01& 1.055E+01& 2.690E+00\\
 3.53& 3.393E+03& 1.395E+00& 7.321E-01& 3.635E-01& 9.865E+00& 3.018E+00\\
 3.63& 4.292E+03& 1.572E+00& 7.354E-01& 3.655E-01& 9.610E+00& 3.292E+00\\
 3.73& 5.429E+03& 1.752E+00& 7.433E-01& 3.694E-01& 9.976E+00& 3.553E+00\\
 3.84& 6.866E+03& 1.920E+00& 7.551E-01& 3.750E-01& 1.113E+01& 3.947E+00\\
 3.94& 8.685E+03& 2.064E+00& 7.705E-01& 3.844E-01& 1.312E+01& 4.689E+00\\
 4.04& 1.099E+04& 2.185E+00& 7.909E-01& 4.056E-01& 1.577E+01& 5.927E+00\\
 4.14& 1.389E+04& 2.294E+00& 8.220E-01& 4.587E-01& 1.869E+01& 7.609E+00\\
 4.24& 1.758E+04& 2.411E+00& 8.747E-01& 5.748E-01& 2.139E+01& 9.464E+00\\
 4.35& 2.223E+04& 2.560E+00& 9.640E-01& 7.800E-01& 2.341E+01& 1.113E+01\\
 4.45& 2.812E+04& 2.759E+00& 1.101E+00& 1.073E+00& 2.448E+01& 1.232E+01\\
 4.55& 3.556E+04& 3.016E+00& 1.285E+00& 1.419E+00& 2.455E+01& 1.288E+01\\
 4.65& 4.498E+04& 3.353E+00& 1.497E+00& 1.762E+00& 2.373E+01& 1.282E+01\\
 4.76& 5.690E+04& 3.841E+00& 1.708E+00& 2.047E+00& 2.222E+01& 1.224E+01\\
 4.86& 7.197E+04& 4.576E+00& 1.886E+00& 2.241E+00& 2.024E+01& 1.130E+01\\
 4.96& 9.103E+04& 5.599E+00& 2.007E+00& 2.331E+00& 1.803E+01& 1.015E+01\\
 5.06& 1.151E+05& 6.818E+00& 2.060E+00& 2.323E+00& 1.574E+01& 8.907E+00\\
 5.16& 1.456E+05& 8.028E+00& 2.045E+00& 2.234E+00& 1.353E+01& 7.674E+00\\
 5.27& 1.842E+05& 8.996E+00& 1.975E+00& 2.088E+00& 1.147E+01& 6.514E+00\\
 5.37& 2.330E+05& 9.559E+00& 1.865E+00& 1.907E+00& 9.610E+00& 5.464E+00\\
 5.47& 2.947E+05& 9.661E+00& 1.733E+00& 1.711E+00& 7.982E+00& 4.541E+00\\
 5.57& 3.728E+05& 9.347E+00& 1.592E+00& 1.516E+00& 6.583E+00& 3.745E+00\\
 5.67& 4.715E+05& 8.718E+00& 1.453E+00& 1.332E+00& 5.398E+00& 3.072E+00\\
 5.78& 5.964E+05& 7.893E+00& 1.324E+00& 1.165E+00& 4.408E+00& 2.509E+00\\
 5.88& 7.543E+05& 6.978E+00& 1.207E+00& 1.018E+00& 3.589E+00& 2.044E+00\\
 5.98& 9.541E+05& 6.058E+00& 1.105E+00& 8.908E-01& 2.917E+00& 1.663E+00\\
 6.08& 1.207E+06& 5.187E+00& 1.017E+00& 7.837E-01& 2.370E+00& 1.352E+00\\
 6.18& 1.526E+06& 4.398E+00& 9.435E-01& 6.943E-01& 1.927E+00& 1.101E+00\\
 6.29& 1.931E+06& 3.705E+00& 8.822E-01& 6.205E-01& 1.570E+00& 8.981E-01\\
 6.39& 2.442E+06& 3.112E+00& 8.317E-01& 5.602E-01& 1.283E+00& 7.353E-01\\
 6.49& 3.089E+06& 2.612E+00& 7.906E-01& 5.113E-01& 1.053E+00& 6.050E-01\\
 6.59& 3.907E+06& 2.198E+00& 7.572E-01& 4.718E-01& 8.700E-01& 5.010E-01\\
 6.69& 4.942E+06& 1.857E+00& 7.303E-01& 4.400E-01& 7.238E-01& 4.181E-01\\
 6.80& 6.251E+06& 1.579E+00& 7.087E-01& 4.146E-01& 6.075E-01& 3.521E-01\\
 6.90& 7.906E+06& 1.354E+00& 6.914E-01& 3.942E-01& 5.151E-01& 2.997E-01\\
 7.00& 1.000E+07& 1.173E+00& 6.776E-01& 3.780E-01& 4.417E-01& 2.580E-01\\
\hline
  logT& T (K)& $\Upsilon$:  1- 6&  1- 7&  1-12&  1-13& 1-18\\
\hline
  2.00& 1.000E+02& 1.307E+00& 4.810E-01& 4.444E-01& 8.154E-01& 9.113E-01\\
  2.10& 1.265E+02& 1.221E+00& 4.936E-01& 4.540E-01& 8.560E-01& 9.043E-01\\
  2.20& 1.600E+02& 1.164E+00& 5.075E-01& 4.647E-01& 9.006E-01& 8.956E-01\\
  2.31& 2.024E+02& 1.136E+00& 5.222E-01& 4.758E-01& 9.425E-01& 8.845E-01\\
  2.41& 2.560E+02& 1.131E+00& 5.374E-01& 4.866E-01& 9.756E-01& 8.705E-01\\
  2.51& 3.237E+02& 1.147E+00& 5.526E-01& 4.963E-01& 9.969E-01& 8.528E-01\\
  2.61& 4.095E+02& 1.175E+00& 5.670E-01& 5.042E-01& 1.008E+00& 8.304E-01\\
  2.71& 5.179E+02& 1.210E+00& 5.797E-01& 5.102E-01& 1.014E+00& 8.023E-01\\
  2.82& 6.551E+02& 1.246E+00& 5.902E-01& 5.195E-01& 1.028E+00& 7.675E-01\\
  2.92& 8.286E+02& 1.285E+00& 5.983E-01& 5.546E-01& 1.077E+00& 7.260E-01\\
\botrule
\end{tabular}
 \\
\end{table}

 \begin{table}
 \noindent{\label{tabl2} Table 3 continues
 }\\
 \footnotesize
 \begin{tabular}{lllllll}
 \toprule
  logT& T (K)& $\Upsilon$:  1- 6&  1- 7&  1-12&  1-13& 1-18\\
 \midrule
  3.02& 1.048E+03& 1.335E+00& 6.038E-01& 6.781E-01& 1.206E+00& 6.795E-01\\
  3.12& 1.326E+03& 1.414E+00& 6.065E-01& 1.021E+00& 1.485E+00& 6.321E-01\\
  3.22& 1.677E+03& 1.533E+00& 6.080E-01& 1.804E+00& 2.005E+00& 5.891E-01\\
  3.33& 2.121E+03& 1.690E+00& 6.163E-01& 3.326E+00& 2.872E+00& 5.558E-01\\
  3.43& 2.683E+03& 1.873E+00& 6.615E-01& 5.914E+00& 4.173E+00& 5.359E-01\\
  3.53& 3.393E+03& 2.079E+00& 8.171E-01& 9.770E+00& 5.920E+00& 5.317E-01\\
  3.63& 4.292E+03& 2.336E+00& 1.199E+00& 1.478E+01& 7.988E+00& 5.465E-01\\
  3.73& 5.429E+03& 2.688E+00& 1.911E+00& 2.039E+01& 1.011E+01& 5.866E-01\\
  3.84& 6.866E+03& 3.169E+00& 2.955E+00& 2.576E+01& 1.198E+01& 6.590E-01\\
  3.94& 8.685E+03& 3.769E+00& 4.202E+00& 3.003E+01& 1.332E+01& 7.674E-01\\
  4.04& 1.099E+04& 4.427E+00& 5.432E+00& 3.264E+01& 1.402E+01& 9.097E-01\\
  4.14& 1.389E+04& 5.047E+00& 6.429E+00& 3.343E+01& 1.406E+01& 1.080E+00\\
  4.24& 1.758E+04& 5.531E+00& 7.061E+00& 3.259E+01& 1.357E+01& 1.273E+00\\
  4.35& 2.223E+04& 5.813E+00& 7.294E+00& 3.053E+01& 1.268E+01& 1.480E+00\\
  4.45& 2.812E+04& 5.870E+00& 7.175E+00& 2.768E+01& 1.154E+01& 1.685E+00\\
  4.55& 3.556E+04& 5.718E+00& 6.789E+00& 2.446E+01& 1.029E+01& 1.865E+00\\
  4.65& 4.498E+04& 5.401E+00& 6.230E+00& 2.117E+01& 9.026E+00& 1.993E+00\\
  4.76& 5.690E+04& 4.969E+00& 5.576E+00& 1.802E+01& 7.809E+00& 2.053E+00\\
  4.86& 7.197E+04& 4.469E+00& 4.890E+00& 1.514E+01& 6.679E+00& 2.038E+00\\
  4.96& 9.103E+04& 3.942E+00& 4.216E+00& 1.258E+01& 5.656E+00& 1.956E+00\\
  5.06& 1.151E+05& 3.420E+00& 3.583E+00& 1.036E+01& 4.749E+00& 1.825E+00\\
  5.16& 1.456E+05& 2.926E+00& 3.009E+00& 8.470E+00& 3.957E+00& 1.663E+00\\
  5.27& 1.842E+05& 2.474E+00& 2.502E+00& 6.885E+00& 3.277E+00& 1.487E+00\\
  5.37& 2.330E+05& 2.072E+00& 2.064E+00& 5.571E+00& 2.700E+00& 1.312E+00\\
  5.47& 2.947E+05& 1.722E+00& 1.691E+00& 4.490E+00& 2.215E+00& 1.145E+00\\
  5.57& 3.728E+05& 1.423E+00& 1.378E+00& 3.608E+00& 1.812E+00& 9.941E-01\\
  5.67& 4.715E+05& 1.170E+00& 1.118E+00& 2.892E+00& 1.479E+00& 8.604E-01\\
  5.78& 5.964E+05& 9.593E-01& 9.047E-01& 2.314E+00& 1.207E+00& 7.448E-01\\
  5.88& 7.543E+05& 7.849E-01& 7.303E-01& 1.849E+00& 9.842E-01& 6.467E-01\\
  5.98& 9.541E+05& 6.418E-01& 5.888E-01& 1.477E+00& 8.038E-01& 5.646E-01\\
  6.08& 1.207E+06& 5.252E-01& 4.744E-01& 1.179E+00& 6.582E-01& 4.965E-01\\
  6.18& 1.526E+06& 4.307E-01& 3.824E-01& 9.408E-01& 5.410E-01& 4.407E-01\\
  6.29& 1.931E+06& 3.544E-01& 3.087E-01& 7.513E-01& 4.471E-01& 3.953E-01\\
  6.39& 2.442E+06& 2.931E-01& 2.497E-01& 6.007E-01& 3.719E-01& 3.585E-01\\
  6.49& 3.089E+06& 2.440E-01& 2.027E-01& 4.809E-01& 3.119E-01& 3.288E-01\\
  6.59& 3.907E+06& 2.047E-01& 1.652E-01& 3.859E-01& 2.641E-01& 3.050E-01\\
  6.69& 4.942E+06& 1.734E-01& 1.354E-01& 3.106E-01& 2.261E-01& 2.859E-01\\
  6.80& 6.251E+06& 1.485E-01& 1.117E-01& 2.508E-01& 1.959E-01& 2.707E-01\\
  6.90& 7.906E+06& 1.287E-01& 9.292E-02& 2.035E-01& 1.719E-01& 2.586E-01\\
  7.00& 1.000E+07& 1.129E-01& 7.802E-02& 1.661E-01& 1.529E-01& 2.490E-01\\
  \botrule
   \end{tabular}
  \end{table}

\subsection{Energy levels and radiative transition parameters of Ca IV}

We present about 93000 radiative transition rates for Ca IV obtained 
from atomic structure calculations in Breit-Pauli approximation
implemented in code SUPERSTRUCTURE. The 13 configurations of Ca IV, 
listed in Computation section, resulted in 387 fine structure energy 
levels, and 93,296 radiative transitions of types allowed (E1) and 
forbidden (E2, E3, M1, M2).

Our calculated energies for Ca~IV have been compared with the measured 
values in Table 1 in the Computation section.
They are in good agreement with the observed energies available at NIST 
table. As seen in Table 1, the differences between calculated and 
measured energies are within 5\% for most of the levels except for the 
highest lying $^4D^o$ and $^4F^o$ states where the differences are 
about 8\%. 

The A-values have been benchmark with limited number of available data.
The present transition probabilities are compared, in Table 4, with the 
only 3 available A-values in the NIST compilation, and with those available
from other sources. Transition $3s^23p^5(^2P^o_{3/2}) \rightarrow 
3s^23p^5(^2P^o_{1/2})$ can be both M1 and E2 types.
The present A-value for the M1 transition, 0.545, is in excellent agreement 
with 0.543 of Naqvi \cite{n51}. The present value is also in good agreement 
with, 0.379 by  Huang et al \cite{h83} given the typical magnitude of 
A-values of the order of 10$^{8}$ s$^{-1}$ or higher for E1 transitions. 
Similarly for the very weak E2 transition, the present $A$-value, 
3.82e-05 sec$^{-1}$ is also in very good agreement with, 1.906E-05 sec$^{-1}$, 
of Huang et al \cite{h83}. For the first dipole allowed transitions
$3s^23p^5(^2P^o_{3/2,1/2}) \rightarrow 3s3p^6(^2S_{1/2})$
present A-values agree very well with those of Wilson et al \cite{w00},
but differ from Huang et al and Varsavsky \cite{v61} who also differ
from each other and from Wilson et al.  As Table 4 shows, the present 
results are in agreement with those of Wilson et al for other transitions.

We are providing two calculated A-values for the dipole allowed 
transitions in order to compare the accuracy, the first A-value has been 
obtained using the experimental transition energy and the second one, 
below it, using the calculated energy. We can see that they themselves 
do not differ significantly from each other. 
Some differences among the results are expected due to use of different
optimization of the configurations included in each calculations and
the accuracy in the methods considered for calculations of the transition
parameters.
  \begin{table}
  \caption{Comparison of present  A-values for {Ca-IV} %obtained using SS 
  with those available in literature. For the E1 transitions, the first 
 A-value from the present work represents use of experimental transition 
 energy while the second one (below it) of calculated energy. K, KP are 
 the initial and final transitional energy level indices (as given in 
 Table 1), $\lambda$ is the transition wavelength in $\AA$ unit. The 
 references are given in superscripts.
 %The A-values are in unit of $sec^{-1}$
  }
    \begin{center}
 \footnotesize
     \label{tab:table1}
     \begin{tabular} {c c c c l l}
\toprule
  K &  KP &$\lambda$  &Transition & A(Present) & A(Others) \\
   & & ($\AA$) & & (sec$^{-1}$) & (sec$^{-1}$) \\
\midrule
    1 & 2 &     & $3s^23p^5(^2P^o_{3/2}) \rightarrow 3s^23p^5(^2P^o_{1/2})$ &
    M1:0.545 & M1:0.541$^{\mbox{\cite{n51}}}$,0.3786$^{\mbox{\cite{h83}}}$\\
 % %
    1 & 2 &     & $3s^23p^5(^2P^o_{3/2}) \rightarrow 3s^23p^5(^2P^o_{1/2})$ &
   E2:3.82e-5   & E2:1.906E-5$^{\mbox{\cite{h83}}}$  \\
    1 & 3 & 656 & $3s^23p^5(^2P^o_{3/2}) \rightarrow 3s3p^6(^2S_{1/2}) $ & 
   5.11E+8  & 7.425E+8$^{\mbox{\cite{w00}}}$,1.20E+10$^{\mbox{\cite{v61}}}$,\\
      &   &     &                    & 4.22E+8   & 1.09e+10$^{\mbox{\cite{h83}}}$\\
    2 & 3 & 669.7 & $3s^23p^5(^2P^o_{1/2})  \rightarrow 3s3p^6(^2S_{1/2})$ &
   2.466E+8 & 3.529E+8$^{\mbox{\cite{w00}}}$, 5.4E+9$^{\mbox{\cite{v61}}}$ \\
      &   &     &                           & 2.01E+8 & \\
    1 & 10 & 454.6 & $3s^23p^5(^2P^o_{3/2}) \rightarrow 3s^23p^43d(^4F_{5/2})$ &
   1.84E+6 & 1.692E+6$^{\mbox{\cite{w00}}}$\\
   1 & 12 & 543 & $3s^23p^5(^2P^o_{3/2}) \rightarrow 3s^23p^43d(^2P_{1/2})$ &
   9.696E+6 & 7.889E+6$^{\mbox{\cite{w00}}}$\\
      &   &     &                           & 9.57E+6 & \\
   2 & 12 & 459.5 & $3s^23p^5(^2P^o_{1/2}) \rightarrow 3s^23p^43d(^2P_{1/2})$ &
   3.930E+7 & 3.503E+7$^{\mbox{\cite{w00}}}$\\
      &   &     &                           & 3.87E+7 & \\
   1 & 14 & 440 & $3s^23p^5(^2P^o_{3/2}) \rightarrow 3s^23p^43d(^4P_{1/2})$ &
   1.350E+7 & 1.243E+7$^{\mbox{\cite{w00}}}$\\
      &   &     &                           & 1.31E+7 & \\
   1 & 15 & 439 & $3s^23p^5(^2P^o_{3/2}) \rightarrow 3s^23p^43d(^4P_{3/2})$ &
   3.90E+6 & 5.367E+6$^{\mbox{\cite{w00}}}$\\
      &   &     &                           & 3.77E+6 & \\
   1 & 16 & 437.3 & $3s^23p^5(^2P^o_{3/2}) \rightarrow 3s^23p^43d(^4P_{5/2})$ &
   2.740E+6 & 2.754E+6$^{\mbox{\cite{w00}}}$\\
      &   &     &                           & 2.66E+6 & \\
   1 & 17 & 437.8 & $3s^23p^5(^2P^o_{3/2}) \rightarrow 3s^23p^43d(^2D_{3/2})$ &
   6.530E+7 & 5.201E+7$^{\mbox{\cite{w00}}}$\\
      &   &     &                           & 6.45E+7 & \\
   1 & 18 & 434.6 & $3s^23p^5(^2P^o_{3/2}) \rightarrow 3s^23p^43d(^2D_{5/2})$ &
   1.920E+8 & 1.576E+8$^{\mbox{\cite{w00}}}$\\
      &   &     &                           & 1.90E+8 & \\
    1 & 27 & 341.3 & $3s^23p^5(^2P^o_{3/2}) \rightarrow 3s^23p^43d(^2S_{1/2})$ &
   6.23E+10 & 4.264E+10$^{\mbox{\cite{w00}}}$,6.543E+10$^{\mbox{\cite{w00}}}$\\
      &   &     &                           & 6.43E+10& \\
\botrule
 \end{tabular}
 \end{center}
 \end{table}

The present work provides lifetimes of all 386 excited levels in a file
that is available electronically at NORAD-Atomic-Data \cite{norad}. 
Lifetime of an excited level can be calculated if the A-values or the
radiative decay rates of the level to the lower levels are known.
Lifetimes are measurable at experimental set-ups. 
Table 5 presents lifetimes of a few levels to illustrate their values.  
For each excited level in the table, the level number, configuration 
number, spectroscopic notation and energy are given. This line is 
followed by the A-values of the level decaying to lower levels. 
The A-values are added and the sum is inverted to obtain the lifetime. 
Levels decaying through the forbidden transitions (E2,E3,M1,M2) have 
longer lifetimes than those through dipole allowed transitions (same
spin E1d and intercombination E1i). No lifetimes for Ca IV levels were
found in literature for comparison. However, their accuracies are related
to those of the A-values which have been discussed for Table 4.

   \begin{table}
   \caption{Lifetimes of a few levels illustrating the complete table of
 lifetimes of all 386 levels. Radiative decay rates of level j to various
 lower levels $j \rightarrow i$ are given. Notation C is for configuration, 
 g for statistical weight factor, lv for level, f for f-value for an E1 
 transition or S-values for E2, E3, M1, M2 transitions, A for A-value and E 
 for transition energy.
   }
     \begin{center}
  \footnotesize
      \label{tab:table1}
      \begin{tabular} {cccc cccc cccc}
 \toprule
 Type & LSi & Ci & gi & lvi & LSj & Ci & gj& lvj& f(E1)/S(E2,& Aji & Eij \\
      &     &    &    &     &     &    &   &    & E3,M1,M2) &(s$^{-1}$)&($\AA$)\\
 \midrule
  \multicolumn{12}{l}{lifetime:~sslevel~j=~  2,~Cf=~1,~$^2$P$^o_{1/2}$~[E= 
 3.019E-02~Ry=~3.3134852E+03 /cm]}\\
   E2& $^2P^o$&1&    4 &  1&      $^2P^o$ &1&   2&   2& 1.54E+00& 5.180E-05& 3.0180E+04\\
   M1& $^2P^o$&1&    4 &  1&      $^2P^o$ &1&   2&   2& 1.33E+00& 6.540E-01& 3.0180E+04\\
  \multicolumn{12}{l}{ 
 Summed~A-values:~Af~(forbidden)= 6.541E-01,~Aa~(allowed)= 0.000E+00 s-1}\\
 \multicolumn{12}{l}{
 Total~Sum(Af+Aa)= Aji~(   2 transitions)~to~the~level=      6.541E-01~s-1}\\
  \multicolumn{12}{l}{ Lifetime (=1/Aji)=                                        1.529E+00~s}\\
  & & & & & & & & & & & \\
  \multicolumn{12}{l}{lifetime:~sslevel~j=~  7,~cf=~3,~$^4$4D$_{1/2}$~[E= 1.826E+00 Ry= 2.0041544E+05 /cm]}\\
  E1i&$^2P^o$& 1&    4&   1&   $^4D$& 3&   2&   7& 1.01E-06& 5.420E+04& 4.9896E+02\\
  E1i&$^2P^o$& 1&    2&   2&   $^4D$& 3&   2&   7& 3.96E-06& 1.030E+05& 5.0735E+02\\
  E2& $^2S$  & 2&    2&   3&   $^4D$& 3&   2&   7& 0.00E+00& 0.000E+00& 1.7417E+03\\
  M1& $^2S$  &2 &   2 &  3 &   $^4D$& 3&   2&   7& 3.18E-07& 8.120E-04& 1.7417E+03\\
  E2& $^4D$  &3 &   6 &  5 &   $^4D$& 3&   2&   7& 1.15E-03& 2.890E-12& 2.0196E+05\\
  E2& $^4D$  &3 &   6 &  5 &   $^4D$& 3&   2&   7& 1.15E-03& 2.890E-12& 2.0196E+05\\
  E2& $^4D$  &3 &   4 &  6 &   $^4D$& 3&   2&   7& 2.69E-03& 1.120E-13& 4.5789E+05\\
  M1& $^4D$  &3 &   4 &  6 &   $^4D$& 3&   2&   7& 5.97E+00& 8.380E-04& 4.5789E+05\\
 \multicolumn{12}{l}{
  Summed~A-values:~Af~(forbidden)= 1.650E-03,~Aa (allowed)= 1.572E+05~s-1}\\
 \multicolumn{12}{l}{
  Total~Sum(Af+Aa)=~ Aji(   8 transitions)~ to~ the~ level=      1.572E+05 s-1}\\
 \multicolumn{12}{l}{
  Lifetime~(=1/Aji)=                                        6.361E-06~s}\\
 \botrule
  \end{tabular}
  \end{center}
  \end{table}

\section{Conclusion}

We have studied collision strength of Ca IV using a 54-levels close 
coupling wave function expansion that corresponds to target ion 
excitations to high lying levels. This ensures inclusion of
converged contributions of resonances generated by all the levels.
We have demonstrated the effect of number of partial waves in the collision
strength $\Omega$ and showed convergence of partial waves contributing
to collision strengths.

Features of $\Omega$ show resonances in the low energy region but they
converge to the background much before reaching the highest 54th excitation 
of Ca IV.
We find that $\Omega$ of the emission line of 3.2 $\mu$m due to collisional
excitation of $^2P^o_{3/2} - ^2P^o_{1/2}$ of ground configuration $3s^23p^5$
has extensive resonances with enhanced background in the low energy region.
This has resulted in a strong effective collision strength $\Upsilon$ with a
peak around 3 $\times 10^5$ K indicating a distinct presence of an emission 
line when the environmental plasma effects are low.
The 3.2 $\mu$m line is within the wavelength range of JWST.

The present $\Omega$ has shown expected features at  high electron energy,
such as, decaying background for the forbidden transitions, slow decay or 
almost constant value for weak dipole transitions and rising trend of 
Coulomb Bethe ln(E) behavior toward a plateau for strong dipole allowed 
transitions. 

We present a set of over 93,000 radiative transitions among 387 energy 
levels with orbitals going up to 5s in Ca IV. Results include lifetimes 
of all 386 excited levels.

The present results are expected to be accurate and large enough for 
the two processes to provide a complete astrophysical modeling for 
all practical purposes.

\backmatter

%\bmhead{Supplementary information}
\bmhead{Data availability}

%If your article has accompanying supplementary file/s please state so here. 

All atomic data for energies, radiative transitions, collisional excitations, 
and effective collision strengths of a set of transition are available 
online at the NORAD-Atomic-Data database at the Ohio State University at:
https://norad.astronomy.osu.edu/

%Authors reporting data from electrophoretic gels and blots should supply the full unprocessed scans for key as part of their Supplementary information. This may be requested by the editorial team/s if it is missing.

%Please refer to Journal-level guidance for any specific requirements.

\bmhead{Acknowledgments}

All computations were carried on the high performance computers of
the Ohio Supercomputer Center. BS acknowledges of IRSIP fellowship
from the Government of Pakistan to carry out the research at the
Ohio State University.

%Acknowledgments are not compulsory. Where included they should be brief. Grant or contribution numbers may be acknowledged.

%Please refer to Journal-level guidance for any specific requirements.

 \section*{Declarations}

Both authors, S.N. Nahar and B. Shafique, contributed equally on the 
contents of the paper. While SNN trained BS, set up the project, wrote 
necessary program, and remained engaged in studying the project, BS picked 
up all aspects of computations, carried out computations, and was engaged 
in the analysis.

\end{document}